\documentstyle[twocolumn,aps,prl,epsf,floats]{revtex}
\begin{document}

\draft
\preprint{}

\twocolumn[\hsize\textwidth\columnwidth\hsize\csname %
@twocolumnfalse\endcsname

\title{Universality of Shot-Noise in Multiterminal Diffusive Conductors}

\author{Eugene V.\ Sukhorukov\cite{Eugene}\cite{inst}
and Daniel Loss\cite{Daniel}}
\address{Department of Physics and Astronomy,
         University of Basel,\\
         Klingelbergstrasse 82,
         CH--4056 Basel, Switzerland}

\date{January 22, 1998}
%\date{\today}
\maketitle
\begin{abstract}
We prove the universality of shot-noise in
multiterminal diffusive conductors of arbitrary shape and dimension
for purely elastic scattering as well as for hot electrons.
Using a Boltzmann-Langevin approach we reduce
the calculation of shot-noise correlators to the solution
of a diffusion equation.  We show that shot-noise in multiterminal
conductors is a non-local quantity and that exchange effects
can occur without quantum phase coherence even at zero electron temperature.
Concrete numbers for shot-noise are given that can be tested experimentally.
\end{abstract}
]
\narrowtext

Shot-noise \cite{Jongrev} induced by
current or voltage fluctuations in electron transport
is a striking
manifestation of charge quantization. It serves
as a sensitive tool to study correlations
in conductors:
while shot-noise assumes a maximum value (with Poissonian distribution)
in the absence of correlations,
it becomes suppressed when correlations set in as e.g.
imposed by the Pauli principle
\cite{Khlus,Lesovik,Yurke,Buttiker1}
or by electron-electron interactions \cite{Nagaev2,Kozub}.

In diffusive mesoscopic
two-terminal conductors where the inelastic scattering lengths
exceed the system
size the shot-noise suppression factor for ``cold'' electrons
(i.e. for vanishing electron temperature)
was predicted
\cite{Beenakker1,Nagaev1,Jong1,Nazarov,Blanter}
to be $1/3$.  The suppression of shot-noise in diffusive conductors
is now experimentally
confirmed \cite{Liefrink,Steinbach,Schoelkopf,Schoenenberger}.
%in particular the value of $1/3$ \cite{Steinbach}.
While some derivations \cite{Beenakker1,Jong1,Nazarov,Blanter}
%of this $1/3$ factor
are based on a scattering matrix approach and thus a priori include
quantum phase coherence, no such effects  are included in the
semiclassical Boltzmann-Langevin equation approach,
which nevertheless leads to the same result \cite{Nagaev1}. However,
while in the quantum approach for a two-terminal conductor
the factor $1/3$ was even shown to be universal \cite{Nazarov},
the semiclassical derivations given so far \cite{Nagaev1,Jongrev}
are restricted to quasi--onedimensional conductors.
For this reason the factor $1/3$
was subsequently challenged
by Landauer as a numerical coincidence \cite{Landauer2}, which
then raises the intriguing question whether phase coherence is relevant
or not for shot-noise in  diffusive conductors.

Motivated by this situation we obtain new results here
that will shed new light
on this question. In particular, we will demonstrate
that the shot-noise suppression
factor of $1/3$ is  {\it universal}
also in the semiclassical Boltzmann-Langevin approach,
in the sense that
it holds for a
multiterminal diffusive conductor of  arbitrary shape and
disorder distribution.
We first prove this for cold electrons and then for the case of
hot electrons where the suppression factor is $\sqrt{3}/4$.
Thereby we extend previous semiclassical
investigations\cite{Nagaev2,Kozub} for
two-terminal conductors
to an arbitrary multiterminal geometry.
This allows us then to compare our semiclassical approach
with the scattering matrix approach
for multiterminal conductors \cite{Buttiker1,Buttiker2,Martin1}, in
particular with some explicit results recently obtained
for diffusive conductors \cite{Blanter}.
While the universality of shot-noise proven here still does not rule out  a
numerical
coincidence \cite{Landauer2} completely,
it certainly makes it less likely and thereby
gives further
support to the suggestion
\cite{Jong2} that phase coherence is not essential for the
suppression of shot-noise in diffusive conductors.

\begin{figure}
  \begin{center}
    \leavevmode
\epsfxsize=7.5cm
\epsffile{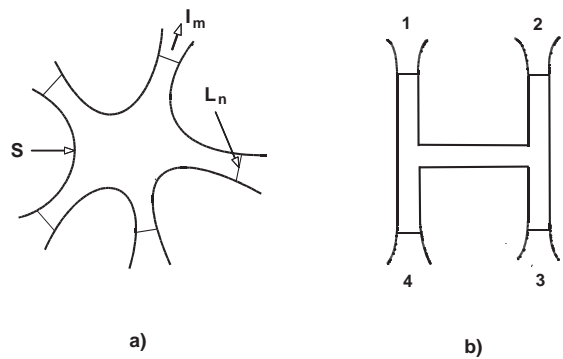}
  \end{center}
\caption{a) Multiterminal diffusive conductor of arbitrary 2D or 3D shape
and with arbitrary impurity distribution.
There are $N$ leads with metallic contacts of area $L_n$,
and $I_m$ is the m-th outgoing current.
$S$ denotes the remaining surface of the conductor where no current 
can pass through. b) H-shaped conductor with four leads of equal
conductances, $G/4$, connected
by a wire in the middle of conductance $G_0$. }
\label{conductor}
\end{figure}

Another remarkable property of shot-noise in mesoscopic
conductors is the exchange effect introduced by
B\"uttiker \cite{Buttiker2}.
Although this effect is generally believed to be phase-sensitive,
we will show that this need not be so. Indeed, for
the particular case of an H-shaped conductor (see Fig. \ref{conductor}b)
we show that exchange effects can be of the same
order as the shot-noise itself even
in the framework of the semiclassical Boltzmann approach.
These exchange effects are shown to come from a non-linear
dependence on the local distribution function. Similarly
we show that the same non-linearities are responsible
for non-local effects
such as the suppression of shot-noise by open leads
even at zero electron temperature.

Finally, we will illustrate the general formalism introduced here
by concrete numbers for various  conductor shapes that are
of direct experimental
interest.

{\em Diffusive conductors.} We consider a multiterminal mesoscopic
diffusive conductor  of arbitrary
2D or 3D shape (see Fig.\ \ref{conductor})
which is
connected to
$N$ perfect metallic contacts of area
$L_n$, $1\leq n \leq N$,
where the voltages $V_n$ or outgoing currents
$I_n$ are measured.
Our goal is to calculate the
multiterminal spectral density $S_{nm}^c$
of current fluctuations $\delta I_n(t)$
at zero frequency $\omega =0$, defined by
\begin{equation}
S^c_{nm}=\int\limits_{-\infty}^{\infty}dt\langle\delta I_n(t)\delta
I_m(0)\rangle .
\label{spectrI}\end{equation}
We start from the (semiclassical)
Boltzmann-Langevin equation for the distribution function
\cite{Kadomtsev,Kogan}. Using standard diffusion approximations
and neglecting the accumulation of charge (we are
interested in $\omega =0$ limit) we get the following equations
\cite{Nagaev2} for the symmetric $f_0(\varepsilon ,{\bf r})$
and antisymmetric
${\bf f}_1(\varepsilon ,{\bf r})$ parts of the average distribution function
$f(\varepsilon ,{\bf r})=f_0(\varepsilon ,{\bf r})+{\bf n}\!\cdot\!{\bf f}_
1(\varepsilon ,{\bf r})$,
%\begin{eqnarray}
%      \nabla [\sigma ({\bf r})\nabla f_0(\varepsilon ,{\bf r})]
%& + & I_{ee}[f]+I_{e-ph}[f]=0,\nonumber\\
%      {\bf f}_1(\varepsilon ,{\bf r})
%& = & v_F\tau_{im}\nabla f_0(\varepsilon,{\bf r}),
%\label{diff}\end{eqnarray}
%
\begin{eqnarray}
      \nabla [\sigma \nabla f_0]
+  I_{ee}[f]
+I_{e-ph}[f]=0,\,\,
      {\bf f}_1=
-v_F\tau_{im}\nabla f_0,
\label{diff}\end{eqnarray}
and for the fluctuations $\delta {\bf j}({\bf r}
,t)$ and $\delta V({\bf r},t)$ of the current density and of the potential,
resp.,
%\begin{equation}
%\delta {\bf j}({\bf r},t)+\sigma ({\bf r})\nabla\delta V({\bf r}\,
%,t)=\delta {\bf j}^s({\bf r},t),
%%\qquad
%\nabla {\bf j}({\bf r},t)=0,
%\label{Langevin}\end{equation}
\begin{equation}
\delta {\bf j}+\sigma \nabla\delta V=\delta {\bf j}^s,\,\,\,
%\qquad
\nabla \cdot \delta {\bf j}=0,
\label{Langevin}\end{equation}
and where the correlator of the Langevin fluctuation
sources is given by
\begin{eqnarray}
      \langle \delta j^s_{\alpha}({\bf r},t)
     \delta j^s_{\beta}({\bf r}^{\prime},t^{\prime})\rangle
& = & \delta_{\alpha\beta}\delta (t-t^{\prime})
\delta ({\bf r}-{\bf r}^{\prime})
\sigma ({\bf r})\Pi ({\bf r}),\nonumber \\
      \Pi ({\bf r})
& = & 2\int d\varepsilon f_0(\varepsilon ,{\bf r})
      [1-f_0(\varepsilon ,{\bf r})].
\label{corr}\end{eqnarray}
Here we introduced the local conductivity
$\sigma ({\bf r})=\gamma e^2\nu_F v_F^2\tau_{im}({\bf r})$
($\gamma_{3D,2D} = 1/3,1/2$),
%for 3D and $1/2$ for 2D cases). ,
electron-electron
$I_{ee}[f]$ and electron-phonon
$I_{e-ph}[f]$ collision integrals,
and used the fact that the electron number is conserved,
%in the scattering process:
$\int d\varepsilon I_{ee}[\delta f]=
\int d\varepsilon I_{e-ph}[\delta f]=0$, where $\delta f$ is the fluctuation
around the  $f$. We also neglected
momentum relaxation due to electron-phonon interaction;
%, i.e. we assume $\tau_{im}\ll\tau_{
%e-ph}.$
but note that we allow for
{\it in}homogeneous disorder, i.e. $\tau_{im}=\tau_{im}({\bf r})$.

Now we specify the boundary conditions to be imposed on
Eqs.\ (\ref{diff},\ref{Langevin}). First, we
assume that
%there are no inelastic processes at the
%open surface $S$ of the conductor (see Fig.\ \ref{conductor}), i.e.
for a given energy there is no current
through the surface $S$. Second, since the contacts with area $L_n$
are perfect conductors the average potential $V$ and its
fluctuations $\delta V$ are independent of position ${\bf r}$ on
$L_n$. Third,
the contacts are assumed to be in thermal equilibrium
with outside reservoirs.
The boundary conditions for (\ref{diff}) and (\ref{Langevin}), resp.,
then read explicitly
\begin{eqnarray}
f_0(\varepsilon ,{\bf r})\left|\right._{L_n}=f_T(\varepsilon -eV_n),\quad
d{\bf s}\!\cdot\!\nabla f_0(\varepsilon ,{\bf r})\left|\right._S=0,
\label{bound1}\\
d{\bf s}\!\cdot\!\delta {\bf j}({\bf r},t)\left|
\right._S=0,\quad\delta V({\bf r}
,t)\left|\right._{L_n}=\delta V_n(t),
\label{bound2}\end{eqnarray}
where $f_T(\varepsilon )$ is the equilibrium distribution function at
temperature $T$, and
$d{\bf s}$ is a vector area element perpendicular to the surface.

We derive now the exact solution of
Eqs.\ (\ref{diff},\ref{Langevin}) with boundary conditions
(\ref{bound1},\ref{bound2}) and use it
to evaluate $S_{nm}^c$ for a multiterminal
conductor of arbitrary shape and distribution of impurities.
For this it is convenient to follow Ref.\onlinecite{Buttiker3}
and to introduce
$N$ characteristic potentials
$\phi_n({\bf r})$ which
satisfy the diffusion equation with associated boundary conditions:
\begin{equation}
\nabla [\sigma \nabla\phi_n]=0, \quad
d{\bf s}\!\cdot\!\nabla\phi_n\left|\right._S=0,\quad
\phi_n\left|\right._{L_m}=\delta_{nm},
\label{potentials}\end{equation}
with $\phi_n\geq 0$ and the sum rule $\sum_{n=1}^N\phi_n=1$. The
potential $V$ (and thus the current ${\bf j}=\sigma \nabla V$)
can then be expressed in terms of $\phi_n$,
$V=\sum_n\phi_n V_n$, and
the conductance then follows as $G_{nm}=-\int_{L_n} d{\bf s}\!\!\ \cdot\!\
\sigma\nabla\phi_m$.
Next we multiply the first part of (\ref{Langevin}) by
$\nabla\phi_n$
and integrate it over the volume.
Subsequent partial integration  of the {\em lhs}\  gives the
solution of (\ref{Langevin}) in terms
of $\phi_n$:
$\delta\tilde {I}_n(t)\equiv\delta I_n(t)-\sum_{m}G_{nm}\delta
V_m(t)=\int d{\bf r}\nabla\phi_n\!\cdot\!\delta {\bf j}^s$.
Using the correlator (\ref{corr})
we obtain a generalized multiterminal spectral density
$S_{nm}=\int_{-\infty}^{\infty}dt\langle\delta\tilde {I}_n(
t)\delta\tilde {I}_m(0)\rangle$:
\begin{equation}
S_{nm}=\int d{\bf r}\sigma\Pi \nabla\phi_n\!\cdot\!\nabla\phi_m,
\label{MSD}\end{equation}
with the properties: $S_{nm}=S_{mn}$ and $\sum_{n}S_{nm}=0$.
Eq.\ (\ref{MSD}) is valid for elastic and inelastic scatterings
and for an arbitrary multiterminal diffusive conductor.
The relation of $S_{nm}$ to the measured noise
is now as follows.
If, say,  the voltages are fixed, then $\delta I_n(t)=\delta
\tilde {I}_n(t)$, and
the matrix $S_{nm}=S_{nm}^c$ is directly measured. On the other
hand, when currents are fixed, $S_{nm}$ can be obtained
from the measured voltage correlator
$S^v_{nm}=\int_{-\infty}^{\infty}dt\langle
\delta V_n(t)\delta V_m(0)\rangle$ by tracing it with conductance matrices:
$S_{nm}=\sum_{n^{\prime}m^{\prime}}G_{nn^{\prime}}G_{mm^{\prime}}S_{
n^{\prime}m^{\prime}}^v$. The physical interpretation of
(\ref{MSD}) becomes now transparent:
$\Pi$ describes the broadening of the distribution
function (effective temperature) that is induced by the voltage applied
to the conductor and $\sigma\Pi$
can thus be thought of as a local noise {\it source}, while $\phi_n$
can be thought of as the {\it probe} of this local noise.
In particular, this means that only $S_{nm}$ is of physical
relevance but not the current or voltage correlators themselves.
We note that (\ref{MSD}) together with
Eqs.\ (\ref{diff},\ref{bound1},\ref{potentials})
can serve as a starting point for
numerical evaluations of $S_{nm}$.
For purely elastic scattering as well as for hot electrons
it is even possible
to get closed analytical expressions for $S_{nm}$ as we will show next.
The physical conditions for the different transport regimes are
discussed in Ref.\ \cite{Nagaev2}.

{\em Elastic scattering: $I_{ee}=I_{e-ph}=0$.}
For simplicity we
work now in the $T=0$ limit (the generalization to
$T>0$ is straightforward \cite{elsewhere}).
First we note that $f_0$ satisfies the diffusion equation
$\nabla [\sigma\nabla f_0]=0$
with boundary conditions (\ref{bound1}). The solution
in terms of $\phi_n$ then is,
$f_0=\sum_{n}\phi_n \theta(\varepsilon -eV_n)$.
Substituting this $f_0$ into $\Pi $ in (\ref{corr})
we obtain
\begin{equation}
\Pi =e\sum\limits_{k,l}\phi_k\phi_l
\left| V_k-V_l\right| ,
\label{temp1}\end{equation}
which in combination with Eq.\ (\ref{MSD}) gives the general expression
for $S_{nm}$ in the case of elastic scattering.

Now we are in the position to generalize the proof of universality
of the $1/3$-suppression of shot-noise
\cite{Beenakker1,Nagaev1,Jong1,Nazarov} to the case of
an arbitrary {\it multiterminal} diffusive conductor.
%2D or 3D geometry with general impurity distribution.
To be specific we choose
$V_n=\delta_{n1}V_1$, i.e. only contact one has a non-vanishing voltage.
Then, using $\sum_n\phi_n=1$,
we get $\Pi =2eV\phi_1[1-\phi_1]$.
Substituting this into
(\ref{MSD}) and integrating by parts we obtain,
$S_{1n}=2eV\oint d{\bf s}\!\cdot\!\nabla\phi_n\sigma
[\phi_1^2/2-\phi_1^3/3]=-{1\over 3}eI_n$. For $n=1$ we
have $S_{11}={1\over 3}eI$, where $I\equiv -I_1>0$ is the
incoming current. Thus, the
factor ${1\over 3}$ is indeed universal in the sense that it does
not depend on the shape of the conductor nor on its impurity distribution.
This generalizes the known universality of a two--terminal conductor
\cite{Nazarov} to a multiterminal geometry.

Next we specialize to two experimentally important cases.
First we consider a multiterminal conductor
of a star  geometry with $N$ long leads
(but with otherwise arbitrary shape)
that join each other at a small crossing
region.
The resistance of this region is assumed to be
much smaller
than the resistance of the leads.
In the second case the contacts are connected through a wide
region,
where again the
resistance of the conductor comes mainly from the regions near
the contacts, while the resistance of the wide region is
negligible. Both shapes are characterized by the requirement
that $w/L\ll 1$, where $w$ and $L$ are the characteristic
sizes of the contact and of the entire conductor, resp.
In both cases the conductor can be divided into $N$ subsections associated
with a particular contact so that the potential $V$ is
approximately constant (for $w/L\ll1$) on the dividing surfaces.
$G_{nm}$ can then be expressed in terms of the conductances
of these subsections (denoted by $G_n$),
\begin{equation}
G_{nm}=(\alpha_m-\delta_{nm})G_n,\quad\alpha_m\equiv G_m/G,
\label{MC2}\end{equation}
where $G\equiv \sum_{n}G_n$.
Applying now similar arguments as above in the proof
of the 1/3-suppression we find\cite{elsewhere}
the explicit expressions
\begin{eqnarray}
      S_{nm}
& = & {1\over 3}e
      \sum\limits_{k}\alpha_n\alpha_k(J_k+J_n)
(\delta_{nm}-\delta_{km}), \nonumber \\
      J_n
& = & \sum\limits_{l} G_l|V_n-V_l|.
\label{MSD2}\end{eqnarray}
We note that this result is valid up to corrections
of order $w/L$ in 3D and for a star geometry in 2D,
and up to corrections of order $[\ln (L/w)]^{-1}$
for wide conductors in 2D\cite{elsewhere}.

We are now in the position to address the issue
of {\em non-locality} of noise in multiterminal conductors.
For this we consider for instance a star geometry
and assume that the
current enters the conductor through
the $n$-th contact, i.e. $I_n=-I$, and leaves it through
the $m$-th contact, i.e. $I_m=I$,
while the other contacts are  open, i.e. $I_k=0$ for $k\neq n,m$.
From (\ref{MC2}) we obtain for the conductance
$G_nG_m/(G_n+G_m)$ (two contacts are in series), and we see that
it does not depend on the other leads, which simply reflects
the {\em local} nature of diffusive transport.
However, contrary to one's first expectation,
this locality does {\em not} carry over to the noise
in general. Indeed, from (\ref{MSD2})
it follows that
$S_{nm}=-{1\over 3}(\alpha_n+\alpha_m)eI$.
The additional
suppression factor $0<\alpha_n+\alpha_m<1$ for $N>2$ reflects the
{\em non-locality} of the current noise.
For instance, for a cross with $N=4$ equivalent leads we have
$\alpha_m=\alpha_n =1/4$, and thus $S_{nm}=-{1\over 6} eI$.
An analogous reduction factor was obtained in
Ref. \onlinecite{Beenakker1} under a different point of view.
Hence, one cannot disregard
open contacts simply because no current is flowing through them;
on the contrary, these open contacts which are still connected to
the reservoir induce  equilibration of the electron gas
and thereby reduce its current noise.
We emphasize that this non-locality is a purely classical
effect in the sense that no quantum phase interference is involved
(phase coherent effects are {\it not} contained
in our Boltzmann approach); the origin of this non-locality
can be traced back to the non-linear dependence
of $\Pi$ on the distribution $f$ in (\ref{corr}).

Next we discuss exchange effects \cite{Buttiker2}
in a four terminal conductor.
According to Blanter and B\"uttiker \cite{Blanter} they can be
probed by measuring $S_{13}$  in three ways:
$V_n=V\delta_{n2}$ (A), $V_n=V\delta_{
n4}$ (B), and
$V_n=V\delta_{n2}+V\delta_{n4}$ (C). Then we take $\Delta S_{13}=S_{1
3}^C-S_{13}^A-S_{13}^B$ as a
measure of the exchange effect.
It comes now as some surprise that in our semiclassical Boltzmann approach
$\Delta S_{13}$ turns out to be non-zero in general
and can even be of the order of the shot-noise itself.
Again, the reason for that is that $\Pi $ is non-linear in $f_0$
(see (\ref{corr})). So, the value $\Pi^C-\Pi^A-\Pi^B$
which enters $\Delta S_{13}$ is not necessarily zero.
Indeed, while exchange effects vanish for cross
shaped conductors (in agreement with \onlinecite{Blanter} up to corrections
of order $w/L$ which are neglected in our  approximation),
it is not so for an H-shaped conductor (see Fig. \ref{conductor}b).
Calculations similar to those leading to
(\ref{MSD2}) give for this case\cite{elsewhere}:
$\Delta S_{13}={1\over {24}}eVG^2G_0/(G+4G_0)^2$,
where  $G_n=G/4$ are the  conductances (all being equal) of the
outer four leads, while the conductance of the  connecting wire
in the middle
is denoted by $G_0$.
This exchange term $\Delta S_{13}$ vanishes
for $G_0\to\infty$, because then the case of a simple cross is
recovered, and also
for $G_0\to 0$, because then the $1$-st and $3$-rd
contacts are  disconnected.
$\Delta S_{13}$ takes on its maximum value
for $G_0=G_n$ and becomes equal to ${1\over {60}}eI^A$,
where $I^A$ is the current
through the $2$-nd contact for case (A).
Although $\Delta S_{13}$ is positive in this example this is not the case in
general \cite{elsewhere}.

In principle, (\ref{MSD2}) and (\ref{MC2}) allow us
to calculate the noise for arbitrary voltages, but
for illustrative purposes we consider again
the simple case of a cross-shaped conductor
with four equivalent leads,
i.e. $\alpha_n=1/4$.
Suppose the voltage is applied to only one contact, say
$V_1=V$,  $V_{n\neq 1}=0$, and $I=-I_1=3I_{n\neq 1}$.
Then, from (\ref{MSD2})
we obtain: $S_{11}={1\over 3}eI$, $S_{12}=S_{13}=S_{14}=-{1\over 9}
eI$, all being in agreement with
the universal $1/3$-suppression proven above. Then,
$S_{22}=S_{33}=S_{44}={2\over 9}eI$, and $S_{23}=S_{24}=S_{34}=-{1\over {
18}}eI$ \cite{comment}.
These numbers seem to be new and it would be interesting
to test them experimentally.

{\em Hot Electrons.}
In this case $I_{ee}\neq 0$, but still $I_{e-ph}=0$, and we assume that
the electron-electron scattering  is sufficiently strong
to cause the heating of
the electrons (i.e. $l_{ee}=\sqrt {D\tau_{ee}}\ll L$,
where $D$ is the diffusion coefficient and $\tau_{ee}$
the electron-electron relaxation time).
The average distribution then assumes the Fermi-Dirac
form:
$f_0(\varepsilon ,{\bf r})=\left\{1+\exp \left[{{\varepsilon -eV({\bf r}
)}\over {T_e({\bf r})}}\right]\right\}^{-1}$, with
$T_e({\bf r})$ being the local electron
temperature. Substituting this $f_0$ into
(\ref{corr}) we immediately get $\Pi =2T_e$.
To find ${T_e({\bf r})}$ explicitly,
we need to solve the diffusion equation (\ref{diff})
subject to the boundary conditions (\ref{bound1}). For this we
integrate these equations (times energy $\varepsilon$) over $\varepsilon$.
$I_{ee}$ then vanishes and we obtain
the diffusion equation $\nabla [\sigma \nabla\Lambda ]=0$,
with boundary conditions
$d{\bf s}\!\cdot\!\nabla\Lambda \left|\right._S=0$,
and $\Lambda \left|\right._{L_n}={1\over 2}(eV_n)^2$, where
$\Lambda ={1\over 6}(\pi T_e)^2+
{1\over 2}(eV)^2$.
Here we used the condition that $T_e\left|\right._{L_n}=T=0$,
which follows from (\ref{bound1}). These equations can be
solved in terms of $\phi_n$, giving
$\Lambda ={1\over 2}e^2\sum_{n}\phi_nV_n^2$.
Using again $V=\sum_n\phi_n V_n$
we obtain
\begin{equation}
\Pi ={{\sqrt 6}\over {\pi}}
e\left[\sum_{n,m}\phi_n\phi_m
(V_n-V_m)^2\right]^{{1\over 2}},
\label{temp2}\end{equation}
which in combination with (\ref{MSD}) gives the general solution
for the case of hot electrons.

Next we prove that the suppression factor
$\sqrt 3/4$ \cite{Nagaev2,Kozub}
for the case of hot electrons in a multiterminal conductor
is also {\it universal}.
As before
%of purely elastic scattering,
we can consider
the case  where the voltage is applied to only one contact:
$V_n=V_1\delta_{n1}$. Then $\Pi={{2\sqrt 3}\over {\pi}}eV_1
\sqrt{\phi_1-\phi_1^2}$. Using the relations
$\sqrt{\phi_1-\phi_1^2}\nabla\phi_1
={1\over 8}\nabla \left\{\arcsin \Phi +\Phi
\sqrt {1-\Phi^2}\right\}$, where
$\Phi =2\phi_1-1$, we transform the volume integral in
(\ref{MSD}) into a surface integral
and obtain $S_{1n}$ in terms of the outgoing currents: $S_{1n}=-{{\sqrt
3}\over 4}eI_n$,
for $n=1,\ldots ,N$.
In terms of the incoming current $I=-I_1$, we get
$S_{11}={{\sqrt 3}\over 4}eI$, which shows that
the $\sqrt 3/4$-factor is indeed universal
\cite{comment2}.

As an illustrative example we consider again
a cross-shaped conductor with four equivalent leads,
$G_n=G/4$, and where we choose
$V_n=V_1\!\delta_{1n}$, $I=-I_1=3I_{n\neq 1}$. We then find
$S_{11}={{\sqrt 3}\over 4}eI $,
%\approx 0.43eI$,
and $S_{
1n}=-{1\over {4\sqrt 3}}eI$,
%\approx -0.14eI$,
for $n\neq 1$,
while
$S_{nn}=\left({{35\sqrt 3}\over {108}}-{2\over {3\pi}}\right)eI$,
%\approx 0.35eI$,
and
$S_{n\neq m}=-\left({{13\sqrt 3}\over {108}}-{1\over {3\pi}}\right)eI$,
%\approx -0.10eI$,
for $n, m\neq 1$.
These new numbers are consistent with the universal factor $\sqrt 3/4$;
they can be generalized to more complicated shapes\cite{elsewhere}.

To conclude, we have proven the universality of shot-noise
in multiterminal diffusive conductors of arbitrary shape
and disorder distribution within a semiclassical
Boltzmann equation approach.
%(for cold and hot electrons.
We have shown that shot-noise is non-local even in classical
diffusive transport and, similarly, that exchange effects
can occur in the absence of phase coherence.
Finally, we have given new suppression factors for shot-noise
in various geometries
which can be tested experimentally.

We would like to thank M.\ B\"uttiker and Ch.\ Sch\"onenberger
for helpful discussions.
This work is supported by
the Swiss National Science
Foundation.


\begin{references}
\bibitem{comment}
Our  $S_{24}$ differs from the one obtained in \onlinecite{Blanter}.
\bibitem{comment2}
Our universality proof
generalizes the discussion of Ref.\onlinecite{Jong2}
which considers a quasi-one-dimensional wire.
\end{references}
\end{document}